\documentstyle[sprocl]{article}

\input{psfig.sty}

\def\Journal#1#2#3#4{{#1} {\bf #2}, #3 (#4)}

\def\NPB{{\em Nucl. Phys.} B}
\def\PLB{{\em Phys. Lett.}  B}
\def\PRL{\em Phys. Rev. Lett.}
\def\PRD{{\em Phys. Rev.} D}
\def\PRC{{\em Phys. Rev.} C}

\def\be{\begin{equation}}
\def\ee{\end{equation}}
\def\bea{\begin{eqnarray}}
\def\eea{\end{eqnarray}}

\begin{document}

\title{HAVING A BLAST WITH EXCITED BARYONS}
\author{Nimai C. Mukhopadhyay and R.M. Davidson} 
\address{Department of Physics, Applied Physics and Astronomy\\ 
Rensselaer Polytechnic Institute, Troy, NY, 12180, U.S.A.}

\maketitle

\section{Introduction}

\noindent
The study of excited baryons provides a window to look at the
chromodynamic structure of hadrons. It is a severe test of our
ability to apply the standard model to hadronic systems, and
eventually to nuclear systems. BLAST can play a small, but special and
significant, role in the valuable kinematic window of
low $W$ and low $Q^2$.\footnote{Invited talk at
the MIT BLAST workshop, presented by NCM}

\section{Low $W$ and Low $Q^2$}

\noindent
Given the energy of the upgraded Bates facility, at the real photon point
we can explore there in
detail the first resonance region, and perhaps the Roper and a bit of
the $N^*$(1535) and $N^*$(1520). How much can be learned about the latter
three resonances will require a careful study of the accelerator and
detector capabilities at their extreme ranges, and therefore lots of
extra planning by the experimentalists as to how use BLAST best in this
difficult domain. Apart from resonance physics, the pion threshold
region is also accessible and of interest as a test of the chiral perturbation
theory (CHPT).

At the real photon point, it will be difficult for BLAST to compete
with facilities like Mainz, GRAAL, LEGS, etc., but perhaps it could
be complementary. It would also provide vital checks on the BLAST
system by comparing with standard results obtained elsewhere. As one
moves away from the real photon point, there are many windows of
opportunity. For example, CHPT calculations of the pion threshold
region are currently believed to valid up to about $Q^2$ of 0.2 GeV$^2$
\cite{ulf}.
Although some data in this region already exist from NIKHEF \cite{ben}
and Mainz \cite{dis},
a thorough study of this region, including polarization measurements,
would be extremely useful. Such measurements would severly test CHPT,
and therefore, QCD itself.

As one moves away from the threshold region and into the first, and
perhaps second resonance regions, the physics becomes of interest to
hadron models, lattice calculations and eventually a test of non-perturbative
QCD \cite{bag,quark,sky,lat}.
Indeed, low $Q^2$ data, up to roughly 0.5 GeV$^2$
are extremely important for testing
nonrelativistic hadron models, since their predictions become
unreliable at large $Q^2$. Even for a relativistic model such
as the bag model \cite{bag}, low Q$^2$ tests are important since
the center of mass corrections are not perfectly under control and
become increasingly important as Q$^2$ increases.

\section{Low Q$^2$: Example from $\Delta$(1232)}

\noindent
The current status of $E2/M1$ at low Q$^2$, shown in Fig.~1, is rather
chaotic. In particular, it appears to be changing from negative to
positive at about 0.15 GeV$^2$, and then becoming negative again at
about 0.4 GeV$^2$. Since as Q$^2$ $\rightarrow$ $\infty$, this ratio
becomes +1, according to the pQCD counting rules \cite{carl},
it must change sign at least one more time. Needless to
say, this structure is very difficult to explain in any hadron model.
It must be pointed out that this ratio has been extracted from a meagre set
of old electroproduction data totally lacking in polarization
observables. In addition, the systematic errors are probably quite
substantial, but are not shown in Fig.~1. \begin{it}Therefore, this is a region
where BLAST can make a big impact.\end{it}

\begin{center}
\begin{figure}[h]
\psfig{figure=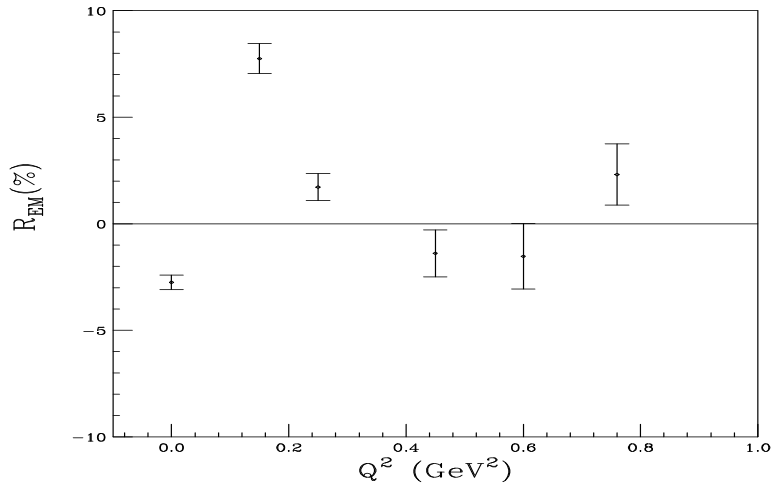,height=2.5in,angle=90,width=4.0in}
\caption{The $E2/M1$ ratio at small $Q^2$ extracted using the
effective Lagrangian approach.\protect\cite{dmw}}
\end{figure}
\end{center}

Regardless of the current status of $E2/M1$ as a function of Q$^2$, it
is clear that the best tests of QCD-inspired hadron models are in the
low Q$^2$ domain where the models are most reliable. At present, the
quark model (in various versions \cite{quark}) is the only practical
description of the resonance region. However, for the $N-\Delta$(1232)
transition, there are predictions from the bag model \cite{bag},
Skryme model \cite{sky}, and pioneering calculations \cite{lat} have
been done on the lattice. Note that while the lattice results are normally
quoted at Q$^2$ = 0, in fact the calculations have been done at nonzero
Q$^2$ and an extrapolation has been made to the real photon point.

\begin{center}
\begin{figure}[h]
\psfig{figure=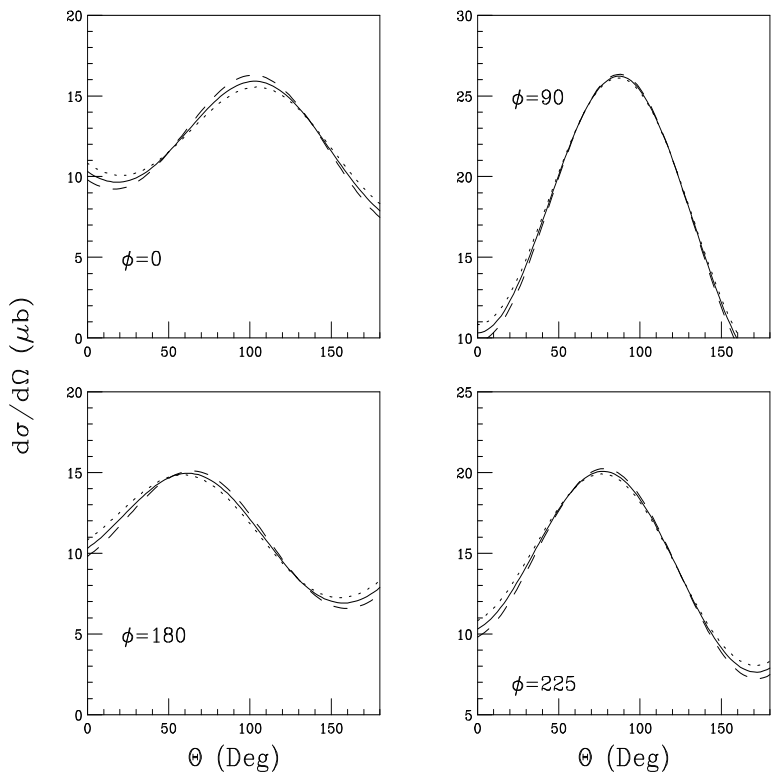,height=4.0in,angle=90,width=4.0in}
\caption{Predictions for the differential cross section based on the
effective Lagrangian approach.\protect\cite{dmw}
In these figures, Q$^2$ = 0.5 GeV$^2$, $W$ = 1.23 GeV and
$\epsilon$ =0.5. The solid curve is with $E2/M1$ = 0\%, the dashed
curve with $E2/M1$ = -10\%, and the dotted curve is with
$E2/M1$ = +10\%.}
\end{figure}
\end{center}

Thus, in the case of the $N-\Delta$ transition, BLAST can provide
precise tests of these QCD-inspired models. In the Roper region, the nature
of this resonance needs to be studied in detail in order to address the
hypothesis \cite{hyb} that it is a hybrid state. Precise experiments
on the $N^*$(1535) would also be of interest \cite{rpi} since its transition
form factor seems to be falling more slowly than a dipole.

Some simple model studies, based on the DMW approach \cite{dmw},
for pion electroproduction at kinematics relelvant to the BLAST
project are shown in Fig.~2. More of these can be obtained from the
authors on request \cite{dav}.

\section{Conclusions}

\noindent
Though the $W$ and Q$^2$ range accessible with BLAST is very narrow, this
is precisely the Q$^2$ range where model calculations are most reliable.
Thus, for the first few resonances, BLAST can provide valuable data to
test CHPT and QCD-inspired models, and ultimately QCD itself. It is of
importance to emphasize that
\begin{it}only very high precision experiments would
be of interest to the physics community, and anything less would not
be useful and would be wasting time and effort.\end{it}

\section*{Acknowledgments}

\noindent
Thanks to Prof.~R.~Milner, Director of the Bates Lab, for his generous support
making NCM's attendance at the BLAST workshop possible. This work
is supported by the U.S.~Dept.~of Energy.

\section*{References}
\bibliographystyle{unsrt}

\end{document}